\documentclass{article}

\usepackage{arxiv}

\usepackage[utf8]{inputenc} 
\usepackage[T1]{fontenc}    
\usepackage{hyperref}       
\usepackage{url}            
\usepackage{booktabs}       
\usepackage{amsfonts}       
\usepackage{nicefrac}       
\usepackage{microtype}      
\usepackage{lipsum}		
\usepackage{graphicx}
\usepackage{natbib}
\usepackage{doi}
\usepackage{color}

\usepackage{amsmath,amssymb,amsfonts}
\usepackage{algorithmic}
\usepackage{textcomp}
\usepackage{import}
\usepackage{multirow} 

\setcitestyle{numbers,square}

\title{CC-DCNet: Dynamic Convolutional Neural Network with Contrastive Constraints for Identifying Lung Cancer Subtypes on Multi-modality Images}

\author{\hspace{1mm}Yuan Jin\\
	Zhejiang Lab, 311121 China\\
	Institute of Computer Graphics and Vision,\\ 
        Graz University of Technology, 8010 Graz, Austria \\
	\And
	\hspace{1mm}Gege Ma \\
	Zhejiang Lab, 311121 China \\
    \And
	\hspace{1mm}Geng Chen \\
	School of Computer Science and Engineering, \\
        Northwestern Polytechnical University,\\
        Xi'an, Shaanxi 710072, China \\
    \And
	\hspace{1mm}Tianling Lyu \\
	Zhejiang Lab, 311121 China \\
    \And
	\hspace{1mm}Jan Egger \\
	Institute of Computer Graphics and Vision, \\
        Graz University of Technology, 8010 Graz, Austria \\
    \And
	\hspace{1mm}Junhui Lyu \\
	Affiliated with the Zhejiang University School of Medicine, \\
        Sir Run Run Shaw Hospital, Hangzhou 310016, China \\
    \And
	\hspace{1mm}Shaoting Zhang \\
        Shanghai Artificial Intelligence Laboratory, Shanghai 200120, China \\
    \And
	\hspace{1mm}Wentao Zhu \\
        Zhejiang Lab, 311121 China \\
}

\renewcommand{\shorttitle}{\textit{arXiv} Template}

\hypersetup{
pdftitle={CC-DCNet: Dynamic Convolutional Neural Network with Contrastive Constraints for Identifying Lung Cancer Subtypes on Multi-modality Images},
pdfsubject={q-bio.NC, q-bio.QM},
pdfauthor={Yuan Jin, Gege Ma, Junhui Lyu, Tianling Lyu, Jan Egger, Geng Chen, and Wentao Zhu},
pdfkeywords={Medical image classification, pathological images, CT, pathological image, neural network},
}

\begin{document}
\maketitle

\begin{abstract}
The accurate diagnosis of pathological subtypes of lung cancer is of paramount importance for follow-up treatments and prognosis managements. Assessment methods utilizing deep learning technologies have introduced novel approaches for clinical diagnosis. However, the majority of existing models rely solely on single-modality image input, leading to limited diagnostic accuracy. To this end, we propose a novel deep learning network designed to accurately classify lung cancer subtype with multi-dimensional and multi-modality images, i.e., CT and pathological images. The strength of the proposed model lies in its ability to dynamically process both paired CT-pathological image sets as well as independent CT image sets, and consequently optimize the pathology-related feature extractions from CT images. This adaptive learning approach enhances the flexibility in processing multi-dimensional and multi-modality datasets and results in performance elevating in the model testing phase. We also develop a contrastive constraint module, which quantitatively maps the cross-modality associations through network training, and thereby helps to explore the ``gold standard'' pathological information from the corresponding CT scans. To evaluate the effectiveness, adaptability, and generalization ability of our model, we conducted extensive experiments on a large-scale multi-center dataset and compared our model with a series of state-of-the-art classification models. The experimental results demonstrated the superiority of our model for lung cancer subtype classification, showcasing significant improvements in accuracy metrics such as ACC, AUC, and F1-score.
\end{abstract}

\keywords{Medical image classification\and pathological images\and CT\and pathological image\and neural network}

\section{Introduction}
\label{sec:introduction}
Lung cancer stands as one of the most prevalent malignant tumors globally, being the foremost cause of cancer-related mortality \cite{in1}. Clinically, the major treatment options are determined on the basis of histopathologic features \cite{in3}. In accordance with the 2015 World Health Organization classification, lung cancer can be classified into two main categories: small-cell lung carcinoma (SCLC) and non-small-cell lung carcinoma (NSCLC) \cite{in4}. NSCLC constitutes around 85\% of lung cancer cases, with lung adenocarcinoma (LUAD) and lung squamous cell carcinoma (LUSC) being the most clinically prevalent histological subtypes \cite{in5}. The differences in treatment approaches exist between LUAD and LUSC, which primarily stem from their distinct molecular and histological characteristics \cite{in6,in7}. For instance, the gene mutations in specific proteins, such as Epidermal Growth Factor Receptor (EGFR), Anaplastic Lymphoma Kinase (ALK) and c-ros oncogene 1 (ROS1) found in LUAD patients can be targeted through precision therapies, while LUSC patients typically lack these molecular targets \cite{in8,in9}, leading the distinct outcomes for the same therapy \cite{in10,in11}. Therefore, developing an effective early screening technique and accurate cancer subtype identification methods becomes critical for further improvements of survival rate for lung cancer patients.

CT has the ability to provide anatomical representations of the human body's internal structures and it can reveal the characteristics of tumors from the macroscopic view, such as size, location, boundaries, morphology, and so forth \cite{in13,in14}. Nevertheless, limited by the relatively low resolution, the accuracy of assessments based on CT images needs further verification, where the atypical clinical manifestations in CT imaging results could pose a challenge in detecting subtle pathological changes through conventional visual assessments \cite{in15}, especially for the complicated cancer pathological subtype identifications \cite{in16}. 

Over the past several years, with the ongoing iterations of emerging information technologies, a plethora of deep learning-based techniques have been extensively proposed and achieved significant breakthroughs \cite{in17}. Benefiting from the characteristics of standardized formats and large datasets, medical imaging has also become a significant application domain for deep learning \cite{in18,in19}. By employing end-to-end deep neural networks to automatically extract and analyse high-throughput features, deep learning models can solve the aforementioned problems in a quantitative way \cite{in20,in21}, and in turn, lead to significant advancements in diagnostic and predictive accuracy. In the context of identifying lung cancer subtypes on CT images, a few explorations of computer-aided systems, leveraging various convolutional neural networks (CNN) and innovative learning strategies, have been conducted \cite{in22,in23,in24,in25}. Despite the achievements, there is still plenty of scope for improvements in the diagnostic accuracy of these automatic analysing techniques \cite{in26}. The limitations of these classification models can be attributed to the limited resolution of CT images as well as the presence of substantial redundant information in the original CT images.

In clinical practice, the one-time result of single-modality examination cannot always provide physicians with sufficient information, thus, the integration of multiple modalities or multiple instances of same modality are frequently required to achieve the complementary of information. In addition to CT tests, a wide array of clinical techniques are also available to aid physicians in diagnosing lung cancer, including pathological examination and other imaging modalities \cite{in27}. While, to acquire the most accurate pathological classification results, pathological examination is highly recommended as the further test. It is regarded as the ``gold standard'' in cancer diagnosis as it can provide highly precise information about pathological types, grading, and staging from the microscopic view \cite{in28,in29}. Consequently, the integration of multi-modality examinations (CT and pathological) can assist physicians in making more accurate diagnoses, which in turn indicates that the deep learning models based on integrating modalities hold the promise of delivering more accurate predictions.

The integration of diverse clinical examinations contributes to providing more comprehensive diagnostic indicators, whereas, it is essential to recognize that the pathological examination is an invasive procedure, which requires tissue specimens through needle biopsy or surgical resection \cite{in30}. Restricted by their physical conditions or the potential risks of complications, patients may not be able to endure the tissue acquisition operations in certain cases \cite{in31}. Thus, the pathological examinations corresponding to certain CT tests are not always obtainable, rendering data missing a common occurrence in clinical applications. As most existing models are designed to handle uniformly-sized image inputs, some studies choose to use only paired data for model training, where the models have limited representation capability and the application value of clinical data may not be fully explored. Additionally, there is some preliminary research that employs generation techniques to complete the missing modality before proceeding with subsequent model training \cite{in32}. Nevertheless, the completion process for datasets with distinguished differences in scale faces great challenges. Recently, to enhance models' representation capability and reduce the computational costs, the concept of dynamic convolution (DC) has been introduced \cite{in33}. DC can adjust the convolution parameters adaptively according to the input images, and use attentions to dynamically invoke and combine convolution kernels to enhance the representation ability of the network \cite{in34}. 

Inspired by such models, we develop a novel lung cancer subtypes classification model that is equipped with dynamic convolutions, which can automatically adjust convolution parameters based on different input combinations, such as the paired multi-modality images or the ones with single-modality, to facilitate diagnosis. Furthermore, to acquire prompts for accurate cancer subtype identification in cases of missing pathological information, we propose contrastive constraints to explore the correlation between paired CT and pathological images, and utilize this as a prior guide for accurate diagnosis in cases of modality absence. We build a multi-scale and multi-center dataset with data from three different tertiary hospitals. With this dataset, our approach outperforms other SOTA methods by a significant margin. 

\section{Methodology}
\label{sec:methodology}
\begin{figure*}[!t]
	\centering
	\includegraphics[width=\linewidth]{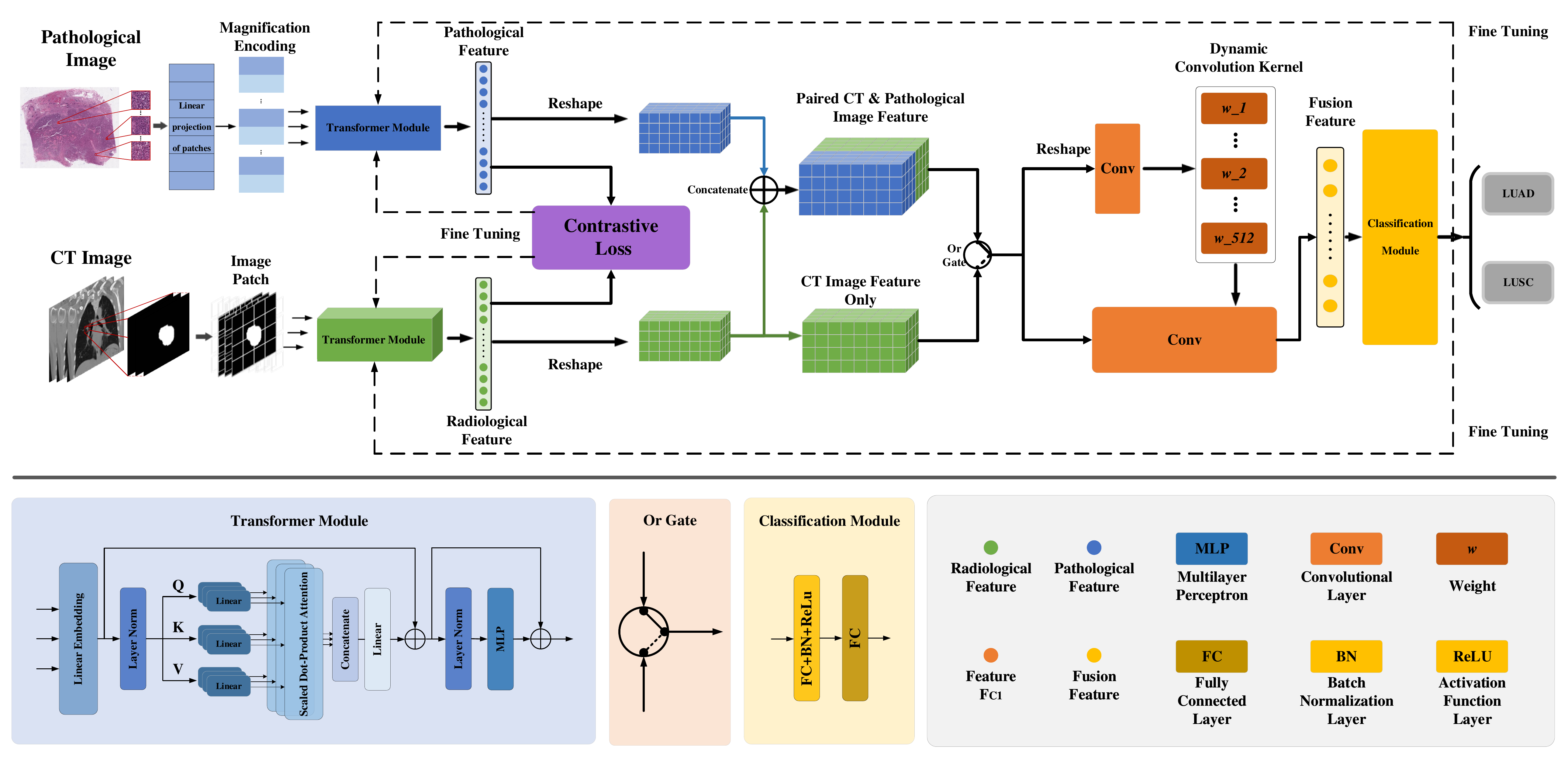}
	\caption{The pipeline of the proposed novel lung cancer subtypes classification model, CC-DCNet.}
	\label{fig:structure}
\end{figure*}

In this section, the proposed CC-DCNet for lung cancer subtype discrimination is introduced, as demonstrated in Fig. \ref{fig:structure}. To address the issues of obtaining relevant pathological information in the absence of pathological images, we propose a novel dynamic convolutional neural network that enables to utilize either paired CT and pathological images or pure CT images as input, and adaptively learns from different input images. Meanwhile, to excavate the cross-modality associations between paired CT and pathological images during training phase, and leverage such associations as priors to influence radiological feature extractions, a contrastive loss is designed to impose the extra constraints. 

\subsection{Dynamic Convolutional Learning}
\label{subsec:dynamic convolutional learning}
In order to enable the model to handle diverse image data, it is necessary for the model to adaptively adjust itself based on input. Most existing static models employ a single convolution kernel to process the image input, resulting in handling the input with the same size only. In this study, inspired by \cite{in33}, we designed a dynamic convolutional network that was constructed with multiple parallel convolution kernels for medical image processing. The parallel convolution kernels can be assembled differently and dynamically regarding to different inputs, and share the same output channels via aggregation. In this way, the model is able to flexibly process various inputs. 

As shown in Fig. \ref{fig:structure}, the pipeline of our proposed model involves three main procedures. Initially, the input images undergo the process of feature extraction, following which the extracted high-level features are concatenated together and subsequently inputted into the parallel convolution kernels to generate predictions for cancer subtypes diagnostic. The detailed procedure is demonstrated as follows.

\subsubsection{Feature Extraction}
The feature extraction part, including radiological feature extractor and pathological feature extractor, plays a crucial role in capturing informative information from input images and performing dimensionality reduction. Notably, the pathological feature extractor is exclusively trained and utilized only when paired CT/pathological data are inputted into the model. In this work, the feature extraction part is equipped with the 3D and 2D vision transformer \cite{math1} network accordingly for CT images and pathological images. 

With regard to the radiological feature extractor, the pre-processed CT images are firstly cropped into patches with an identical size of 112$\times$112$\times$112, and subsequently fed into the Transformer network module. To empower the model to focus on the information within these patches from different sub-spaces, we incorporate the multi-head self-attention mechanism into the network architecture. This mechanism allows the model to jointly capture and learn relevant features from the patches, enhancing its ability to discern important patterns and relationships across the input data. Building upon the 3D vision transformer module, we can extract radiological features $\mathbf{X_r}\in \mathbb{R}^N$ from the original CT images, in this study $N$ is set to 1024. 

With respect to pathological feature extractor, we employed the 2D vision Transformer module to process the high resolution image patches, utilizing operations similar to those in the radiological feature extraction part. Interestingly, our observations indicated that positional information played a less crucial role in the context of our randomly sampled patches sourced from the same Whole Slide Imaging (WSI). Consequently, the incorporation of advanced positional encoding did not yield a substantial improvement in performance. Thus, to augment the model's receptive field for capturing pathological features at varying magnification rates, we replaced the position code with the magnification rate. The desired high-level pathological features are represented as $\mathbf{X_p}\in \mathbb{R}^N$.

\subsubsection{Feature Concatenation}
After the feature extraction parts, both CT and pathological features of size $N$ are obtained. The CT and pathological features are concatenated subsequently and reconstructed into $\mathbf{\Tilde{X}}\in \mathbb{R}^{H\times L\times D}$. It is worth noting that the pathological features are not always available, given the scenarios of both paired CT/pathological images and pure CT images are model inputs. In this study, the size of $\mathbf{\Tilde{X}}$ with paired CT/pathological inputs is set to $16\times16\times8$ and the one with only CT inputs is set to $16\times8\times8$.

\subsubsection{Dynamic Convolution}
The goal of dynamic convolution is to turn input $\mathbf{\Tilde{X}}$ with different sizes into output $\mathbf{Z}\in \mathbb{R}^M$ with the same size. In this study, $M$ is set to 512. Let $x$ and $z$ denote one element of $\mathbf{\Tilde{X}}$ and $\mathbf{Z}$, respectively. With no padding and stride of the same size with $\mathbf{\Tilde{X}}$, the dynamic convolution is processed as follows:

\begin{equation}
\begin{aligned}
z_m&=\sum_{h=0}^{H}\sum_{l=0}^{L}\sum_{d=0}^{D}w_{h,l,d,m}\times x_{H-h-1,L-l-1,D-d-1}
\end{aligned}
\label{eq:dynamic convolution1}
\end{equation}
where $w$ denotes one element of the weight $\mathbf{\Tilde{W}}\in \mathbb{R}^{H\times L\times D\times M}$, which varies in response to $\mathbf{\Tilde{X}}$ and is generated by convolving $\mathbf{\Tilde{X}}$ with $M$ convolution kernels based on same padding \cite{math5}. 

Based on the above derivation, we implement the dynamic convolution module, where the concatenated feature $\mathbf{\Tilde{X}}$ is passed through the Convolution Kernel Generation module (shown in Fig. \ref{fig:structure}). This module performs convolution with an output channel of 512 and a convolution kernel size of $3\times3\times3$, with padding 1 and stride 1, and generates the weight $\mathbf{\Tilde{W}}$ of size $16\times 16\times 8\times 512$.

The feature $\mathbf{\Tilde{X}}$ is then convolved with $\mathbf{\Tilde{W}}$ through a 3D convolution in Eq. \ref{eq:dynamic convolution1}, that has an output channel of 512 and a kernel size of $16\times16\times8$, with no padding and a stride of $16\times16\times8$. This operation produced a 512-dimensional output feature $\mathbf{Z}$, which is finally fed into a fully connected layer to obtain the classification result.

Furthermore, in the absence of pathological features, the width size is adjusted from 16 to 8 in the aforementioned process, while all other steps remain unchanged. This approach still ensures the attainment of the final classification result. This adaptive method facilitates dynamic convolution with varying kernel sizes tailored to different input sizes, accommodating scenarios both with and without pathological features.

\subsection{Contrastive Loss Function}
Numerous studies have demonstrated that the cross-scale correlations between pathological images and CT images for the same cancer subtype can be found \cite{math2,math3}. Therefore, in this work, we designed a comprehensive set of contrastive loss constraints $\mathcal{L}_{\text{contrast}}$ within the proposed dynamic convolutional network, encompassing the contrastive loss for pathological subtypes $\mathcal{L}_{\text{type}}$ and the contrastive loss for cross-modality correlations $\mathcal{L}_{\text{correlation}}$. This contrastive loss set is applied only when paired CT and pathological images are inputted, it can be defined as:
\begin{equation}
\begin{aligned}
{\mathcal{L}_{\text{contrast}}}=\mathcal{L}_{\text{type}}+\lambda_{p}\mathcal{L}_{\text{correlation}}
\end{aligned}
\label{eq:loss_contrastive}
\end{equation}
where $\lambda_{p}$ is the tuning parameter to balance the contributions of these two losses. 
\begin{figure}[!t]
	\centering
	\includegraphics[width=0.85\linewidth]{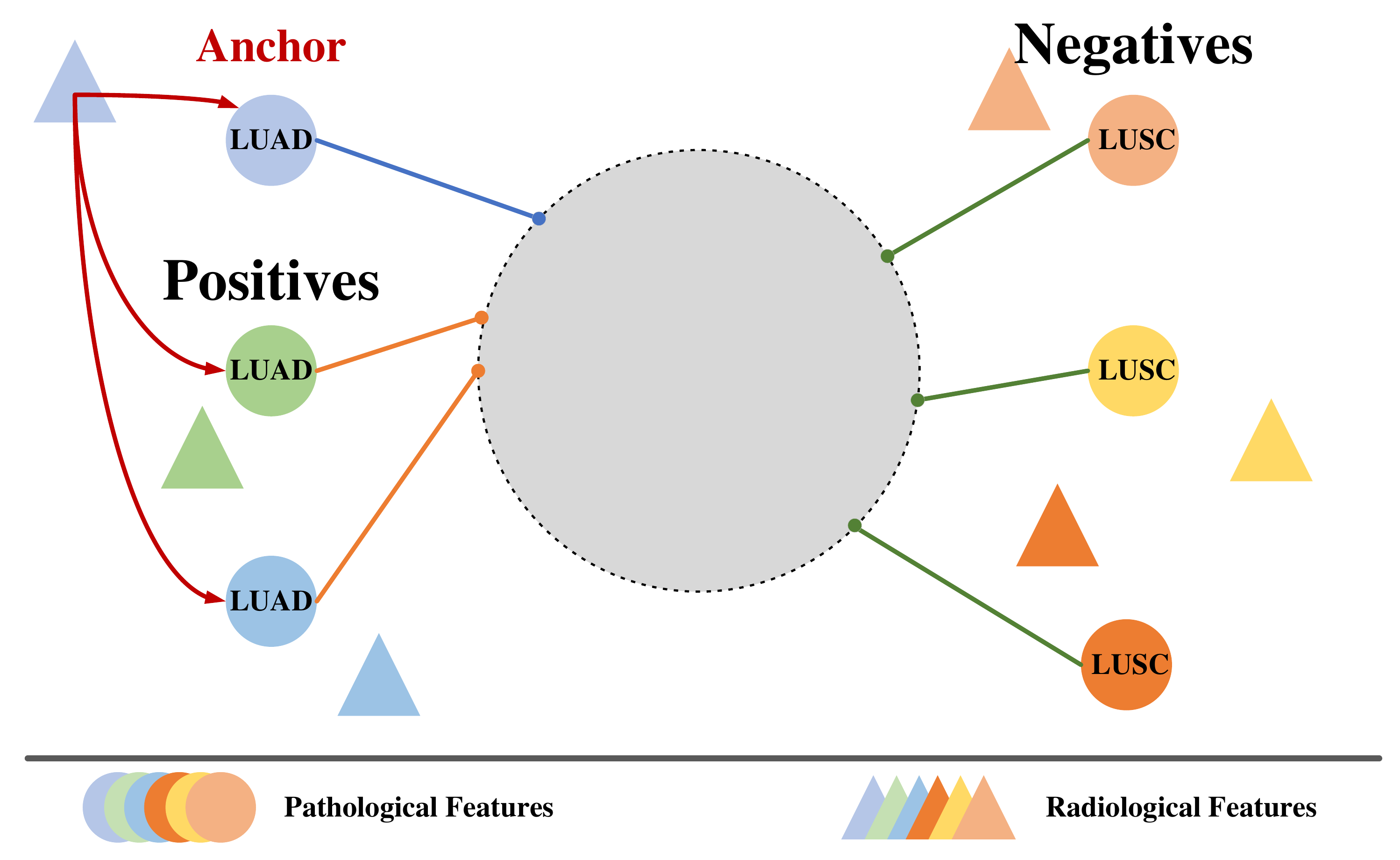}
	\caption{The diagram of contrastive Loss.}
	\label{fig:ContrustiveLoss}
\end{figure}

The role of contrastive loss is to map similar features into close regions, while mapping dissimilar ones to distant regions, thereby assisting the network in learning discriminative feature representations and bringing similar features closer in the feature space \cite{math4}. In our design, as shown in Fig. \ref{fig:ContrustiveLoss}, the contrastive loss for pathological subtypes $\mathcal{L}_{\text{type}}$ is employed to empower the network to distinguish between different cancer subtypes, meanwhile, the contrastive loss for cross-modality correlations $\mathcal{L}_{\text{correlation}}$ is used to enforce the network to extract highly correlated features from both CT and pathological images. In this way, with the guidance of gold standard pathological images, our approach enables the radiological feature extractor to capture features that are more relevant to pathological information, thereby enhancing the subsequent diagnostic accuracy. The loss can be represented as follows:
\begin{equation}
\begin{aligned}
{\mathcal{L}_{\text{type}}}=\sum_{i\in I}\frac{-1}{|P(i)|}\sum_{p\in P(i)}\log\frac{\exp(k_i\cdot k_p/\tau)}{\sum_{n\in N(i)}\exp(k_i\cdot k_n/\tau)}
\end{aligned}
\label{eq:loss_type}
\end{equation}
where $P(i)$ is the set of indexes of all samples in the batch, which have the same cancer subtype as $i$, $k_i$ refers to the feature from the whole set, $k_p$ represents another feature with the same cancer subtype,  $n\in N(i)$ refers to all samples from the whole dataset, the $\cdot$ symbol refers to the inner product, $\tau$ is a scalar temperature parameter.
\begin{equation}
\begin{aligned}
{\mathcal{L}_{\text{correlation}}}=-\sum_{j\in I}\log\frac{\exp(k_j\cdot k_+/\tau)}{\sum_{a\in A(j)}\exp(k_j\cdot k_a/\tau)}
\end{aligned}
\label{eq:loss_patient}
\end{equation}
where $k_j$ refers to the feature from an original CT image, $k_+$ represents its corresponding pathological image feature, $a\in A(j)$ refers to all samples from the paired CT/pathological dataset, the $\cdot$ symbol refers to the inner product, $\tau$ is also a scalar temperature parameter.

\subsection{The Overall Loss Function}
The hybrid feature $F_{C2}$ acquired from the dynamic convolution module will give an ultimate classification outcome, where the loss can be calculated to update the entire model. In addition, the applied contrastive loss can also improve the model's ability in cancer subtypes discrimination and correlated multi-modality feature extraction. Therefore, in order to more effectively incorporate the pathological features as the benchmark to enhance the entire model's classification accuracy, even when CT images are utilized solely as input, we designed a total loss function that is outlined as follows:
\begin{equation}
\begin{aligned}
&{\mathcal{L}_{\text{total}}}=\mathcal{L}_{\text{class}}+\alpha \lambda_{c}\mathcal{L}_{\text{contrast}},\\
&\quad\left\{\begin{array}{l}
\alpha=1, \text { when inputs are paired CT/pathological images} \\
\alpha=0, \text { when inputs are standalone CT images}
\end{array}\right.
\end{aligned}
\label{eq:loss_final}
\end{equation}
where $\mathcal{L}_{\text{class}}$ and $\mathcal{L}_{\text{contrast}}$ are the representative signs of the classification loss and contrastive loss, respectively; $\alpha$ is the parameter of 0 or 1; $\lambda_{c}$ is the tuning parameter to balance the contributions of classification loss and contrastive loss.

For the classification loss, we applied the cross entropy loss as defined in Eq. \eqref{eq:CEL},
\begin{equation}
\begin{aligned}
\operatorname {\mathcal{L}_{\text{class}}} =-\frac{1}{N} \sum_{t} [y_{t} \cdot \log \left(p_{t}\right)+(1-y_{t}) \cdot \log \left(1-p_{t}\right)],
\end{aligned}
\label{eq:CEL}
\end{equation}
where ${\mathcal{L}_{\text{class}}}$ refers to the loss, $N$ refers to the total number of training samples, and $t$ refers to the $t^{th}$ sample. $y_{t}$ is the label of the $t^{th}$ sample, and it is a sign function, which equals one if the label is the same as LUSC and zero otherwise. $p_{t}$ is the probability that the prediction of the $t^{th}$ sample belongs to class LUSC.

\subsection{Evaluation Metrics}
\label{subsec:metrics}
To evaluate the performance of our proposed CC-DCNet, the area under the curve (AUC) was utilized as the metric for binary classification. Moreover, accuracy (ACC) as well as F1-score were calculated to find the optimized threshold. Each of these metrics encompasses a value that falls within the range of 0 to 1, wherein a higher value signifies an enhanced performance of the model.

\section{Experimental Setup}
\label{sec:results}
\begin{table*}[!t]\small
\caption{Clinical information of the multi-center patient cohort}
\label{tab:databases}
\centering
\renewcommand{\arraystretch}{1.1}
\setlength{\tabcolsep}{2.4mm}{
\small
\tabcolsep=0.08cm
\begin{tabular}{ccccccccc}
 \hline \hline
\multicolumn{3}{c||}{\begin{tabular}[c]{@{}c@{}}Hospital A\\ (Sir Run Run Shaw Hospital, \\Zhejiang University School of Medicine )\end{tabular}} 
& \multicolumn{3}{c||}{\begin{tabular}[c]{@{}c@{}}Hospital B\\ (Affiliated Dongyang Hospital \\ of Wenzhou Medical University)\end{tabular}}  
& \multicolumn{3}{c}{\begin{tabular}[c]{@{}c@{}}Hospital C\\ (Cancer Hospital of The University \\of Chinese Academy of Sciences)\end{tabular}}   \\ \hline
   
\multicolumn{1}{c}{\begin{tabular}[c]{@{}c@{}}Patient\\ (n=520)\end{tabular}}        
& \multicolumn{1}{c}{\begin{tabular}[c]{@{}c@{}}LUAD\\ (n=293)\end{tabular}} 
& \multicolumn{1}{c||}{\begin{tabular}[c]{@{}c@{}}LUSC\\ (n=227)\end{tabular}} 
& \multicolumn{1}{c}{\begin{tabular}[c]{@{}c@{}}Patient\\ (n=191)\end{tabular}}        
& \multicolumn{1}{c}{\begin{tabular}[c]{@{}c@{}}LUAD\\ (n=106)\end{tabular}}        
& \multicolumn{1}{c||}{\begin{tabular}[c]{@{}c@{}}LUSC\\ (n=85)\end{tabular}}    
& \multicolumn{1}{c}{\begin{tabular}[c]{@{}c@{}}Patient\\ (n=435)\end{tabular}}        
& \multicolumn{1}{c}{\begin{tabular}[c]{@{}c@{}}LUAD\\ (n=235)\end{tabular}}        
& \begin{tabular}[c]{@{}c@{}}LUSC\\ (n=200)\end{tabular}       \\ 

\multicolumn{1}{c}{\begin{tabular}[c]{@{}c@{}}Paired\\ (n=320)\end{tabular}}        
& \multicolumn{1}{c}{\begin{tabular}[c]{@{}c@{}}LUAD\\ (n=185)\end{tabular}} 
& \multicolumn{1}{c||}{\begin{tabular}[c]{@{}c@{}}LUSC\\ (n=135)\end{tabular}} 
& \multicolumn{1}{c}{\begin{tabular}[c]{@{}c@{}}Paired\\ (n=191)\end{tabular}}        
& \multicolumn{1}{c}{\begin{tabular}[c]{@{}c@{}}LUAD\\ (n=106)\end{tabular}}        
& \multicolumn{1}{c||}{\begin{tabular}[c]{@{}c@{}}LUSC\\ (n=85)\end{tabular}}    
& \multicolumn{1}{c}{\begin{tabular}[c]{@{}c@{}}Paired\\ (n=0)\end{tabular}}        
& \multicolumn{1}{c}{\begin{tabular}[c]{@{}c@{}}LUAD\\ (n=0)\end{tabular}}        
& \begin{tabular}[c]{@{}c@{}}LUSC\\ (n=0)\end{tabular}       \\ 

\multicolumn{1}{c}{\begin{tabular}[c]{@{}c@{}}CT only\\ (n=200)\end{tabular}}        
& \multicolumn{1}{c}{\begin{tabular}[c]{@{}c@{}}LUAD\\ (n=108)\end{tabular}} 
& \multicolumn{1}{c||}{\begin{tabular}[c]{@{}c@{}}LUSC\\ (n=92)\end{tabular}} 
& \multicolumn{1}{c}{\begin{tabular}[c]{@{}c@{}}CT only\\ (n=0)\end{tabular}}        
& \multicolumn{1}{c}{\begin{tabular}[c]{@{}c@{}}LUAD\\ (n=0)\end{tabular}}        
& \multicolumn{1}{c||}{\begin{tabular}[c]{@{}c@{}}LUSC\\ (n=0)\end{tabular}}    
& \multicolumn{1}{c}{\begin{tabular}[c]{@{}c@{}}CT only\\ (n=435)\end{tabular}}        
& \multicolumn{1}{c}{\begin{tabular}[c]{@{}c@{}}LUAD\\ (n=235)\end{tabular}}        
& \begin{tabular}[c]{@{}c@{}}LUSC\\ (n=200)\end{tabular}       \\ \hline

\multicolumn{1}{c}{Age (year)}      
& \multicolumn{1}{c}{60.89$\pm$10.41}   
& \multicolumn{1}{c||}{65.34$\pm$8.72}   
&\multicolumn{1}{c}{Age (year)}      
& \multicolumn{1}{c}{60.43$\pm$11.03} 
& \multicolumn{1}{c||}{66.71$\pm$6.83} 
& \multicolumn{1}{c}{Age (year)}      
& \multicolumn{1}{c}{62.19$\pm$10.50} 
& 65.37$\pm$8.96 \\ 

\multicolumn{1}{c}{Male}   
& \multicolumn{1}{c}{108}   
& \multicolumn{1}{c||}{227}  
& \multicolumn{1}{c}{Male}   
& \multicolumn{1}{c}{35}          
& \multicolumn{1}{c||}{85}         
& \multicolumn{1}{c}{Male}   
& \multicolumn{1}{c}{95}          
& 172         \\ 

\multicolumn{1}{c}{Female} 
& \multicolumn{1}{c}{176}   
& \multicolumn{1}{c||}{9}  
& \multicolumn{1}{c}{Female} 
& \multicolumn{1}{c}{71}          
& \multicolumn{1}{c||}{0}          
& \multicolumn{1}{c}{Female} 
& \multicolumn{1}{c}{160}          
& 8          \\ \hline  \hline
\end{tabular}}
\end{table*}
\subsection{Datasets and Preprocessing}
This study was conducted using a multi-center dataset formed by three distinct hospitals: Sir Run Run Shaw Hospital, Zhejiang University School of Medicine (referred to as Hospital A); the Affiliated Dongyang Hospital of Wenzhou Medical University (referred to as Hospital B); and the Cancer Hospital of The University of Chinese Academy of Sciences (referred to as Hospital C). Table \ref{tab:databases} presents the detailed demographic information of all patients participating in this study. A total of 520 LUAD/LUSC patients from Hospital A, who had undergone either CT scanning alone or a combination of CT scanning and biopsy/surgical specimen examinations, were encompassed within this study. Furthermore, 191 and 435 patients with LUAD/LUSC diagnoses from Hospital B and Hospital C were also included in the study. The data retrospectively collected from Hospital A, encompassing both paired CT/pathological images and pure CT images were employed for model training and testing, while the data obtained from Hospitals B and C, featuring a smaller number of paired data and a larger pool of pure CT images, were used to assess the stability and generalization ability of the well-trained network. Prior to training and testing the proposed model, it is necessary to preprocess both the original pathological and CT images, and the following are the explanations of the relevant preprocessing procedures.

\subsubsection{Pathological Image Preprocessing}
\label{subsubsec:pathological pre-processing}
The pathological images employed in this study were in the form of whole slide images (WSIs), which represent digitized renditions of glass slides achieved through dedicated slide scanners \cite{es1}. The dimensions of all WSIs spanned an average width of 77460$\pm$14662 pixels and an average height of 59317$\pm$11014 pixels. Confronting the challenge posed by these extensive image sizes, a transformer network was introduced to handle patches extracted from the original WSI images, encompassing varying magnification levels. This approach effectively mitigated the intricacies of working with these sizable images. Furthermore, to ensure the fidelity and consistency of the feature information, two essential steps were undertaken prior to patch extraction: color normalization of all WSIs and delineation of the region of interest (ROI). To counteract the impact of color deviations, this study incorporated a technique known as structure-preserving color normalization (SPCN) \cite{es2,es3}. Then the suspected cancerous regions within all WSIs were meticulously annotated by three pathologists utilizing the Automated Slide Analysis Platform (ASAP) \cite{es4}. This rigorous process was aimed at solely focusing on features pertinent to cancer. Subsequent to the delineation, patches were exclusively cropped within the defined ROI. Following this, the original WSIs were cropped at four distinct magnification levels, namely 10\textbf{X}, 20\textbf{X}, 40\textbf{X}, and 100\textbf{X}, while being standardized to a uniform patch size of 560$\times$560 pixels, where \textbf{X} means times.

\subsubsection{CT Image Preprocessing}
\label{subsubsec:CT pre-processing}
To ensure comprehensive coverage of malignancies across all patient cases, all of the CT raw data underwent reconstruction with the same parameters, maintaining an in-plane resolution of 1.0mm and a slice thickness of 1.0mm. Besides, tumor segmentation of the multi-center CT images was executed by a radiology specialist employing the medical image processing software ITK-SNAP 3.8 \cite{es5}. To encompass all cancer-related features, the cancer masks were automatically dilated by three voxels as part of the region delineation process. Subsequent to this, a final Volume of Interest (VOI) of fixed dimensions, specifically 256$\times$256$\times$128, was cropped for each CT image. Furthermore, to ensure standardized data representation, the values of all generated patches were normalized to reside within the range of zero to one.

\subsection{Implementation Details}
Throughout the training process, the complete dataset underwent division into two parts: a training dataset and a testing dataset, distributed in an 80\% to 20\% ratio. Subsequent to the implementation of a five-fold cross-validation technique, the patches associated with different subjects and their corresponding pathological labels were systematically utilized to train the model. The total loss function played a pivotal role in iteratively updating the model parameters throughout the entire training phase.

In this work, our proposed CC-DCNet and the other comparison methods were implemented in PyTorch with CUDA 10.2 toolkit and CUDA Deep Neural Network (cuDNN) 8.0.2 on the Ubuntu 18.04 operating system. All experiments involving training, validating, and testing were conducted on a Tesla V100-SXM2 graphics card. During the training phase, the parameters are set as 4 for batch size and 0.0001 for initial learning rate. For each approach, 400 epochs were executed, and the adaptive moment estimation (Adam) optimizer was utilized to optimize parameter updates. 

\begin{table*}[!t]
\centering
\caption{The comparisons between the proposed model and other SOTAs in terms of ACC(\%), AUC(\%) and F1-Score(\%), tested with independent CT inputs.}
\renewcommand{\arraystretch}{1.5}
\setlength{\tabcolsep}{6.5pt}
\small
\tabcolsep=0.08cm
\begin{tabular}{cccccccccc}
\hline
Metrics & ResNet-18 & ResNet-50 & ViT & T2T & Swin &SwitchNet &RAFENet&CARL&CC-DCNet  \\ \hline 
ACC &79.31$\pm$4.24  &79.51$\pm$4.03 &80.33$\pm$2.99 &80.58$\pm$2.18 &80.73$\pm$3.06 &81.70$\pm$3.01 &82.67$\pm$3.89 &83.14$\pm$2.94 &\textcolor[RGB]{255,0,0}{86.69$\pm$3.25}\\
AUC &84.31$\pm$4.28  &84.33$\pm$3.48 &84.81$\pm$4.10 &84.83$\pm$2.10  &85.19$\pm$2.46 &86.09$\pm$3.26 &87.05$\pm$2.51 &87.61$\pm$3.06 &\textcolor[RGB]{255,0,0}{90.75$\pm$3.46}\\
F1-Score &79.04$\pm$4.49  &79.50$\pm$3.82 &79.92$\pm$2.64 &79.93$\pm$2.40 &79.98$\pm$3.04 &81.64$\pm$2.52 &82.05$\pm$3.51 &82.55$\pm$2.70 &\textcolor[RGB]{255,0,0}{85.96$\pm$3.19}\\ \hline 
\end{tabular}
\label{tab:comparisonstudy1}
\end{table*}


\begin{table}[!t]
\centering
\caption{The comparisons between the proposed model and other SOTAs in terms of ACC(\%), AUC(\%) and F1-Score(\%), tested with paired CT/pathological inputs. The results of RAFENet and CARL are omitted, as these models were specifically designed to process single-modality images.}
\renewcommand{\arraystretch}{1.3}
\tabcolsep=10pt
\begin{tabular}{cccc}
\hline
Metrics & ACC& AUC &F1-Score  \\ \hline 
ResNet-18 &93.75$\pm$2.66 &95.69$\pm$2.68 &93.67$\pm$2.44     \\
ResNet-50&93.97$\pm$2.89   &95.96$\pm$3.01 &93.82$\pm$3.13    \\
ViT &94.39$\pm$2.74 &96.36$\pm$2.66  &93.97$\pm$3.03    \\ 
T2T&94.27$\pm$2.77&96.19$\pm$2.69&93.43$\pm$3.73\\
Swin&94.53$\pm$3.10&96.39$\pm$2.37&94.01$\pm$3.06\\
SwitchNet&94.65$\pm$2.36&96.53$\pm$2.07&94.61$\pm$2.98\\
CC-DCNet&\textcolor[RGB]{255,0,0}{96.28$\pm$2.56}&\textcolor[RGB]{255,0,0}{97.53$\pm$2.13}&\textcolor[RGB]{255,0,0}{96.04$\pm$3.33}\\ \hline 
\end{tabular}
\label{tab:comparisonstudy2}
\end{table}

\section{Experimental Results}
\label{sec:results}
In this section, we first evaluated the performance of proposed CC-DCNet through a comparative study against other state-of-the-art (SOTA) models. We then investigated the contributions of key components of our model by conducting the ablation tests regarding contrastive constraints and additional independent CT dataset, followed by an impact analysis of utilizing different SOTAs for feature extraction. Lastly, the model underwent testing on a multi-center external dataset to evaluate its generalization ability.
\subsection{Comparison with the State-of-the-Arts}
Given that our research incorporates inputs with varying sizes, including both paired CT/pathological images and independent CT images, the comparison study was conducted in two parts: one to compare with SOTAs on CT dataset, and the other with SOTAs on paired CT/pathological datasets.

\begin{figure}[!h]
	\centering
	\includegraphics[width=\linewidth]{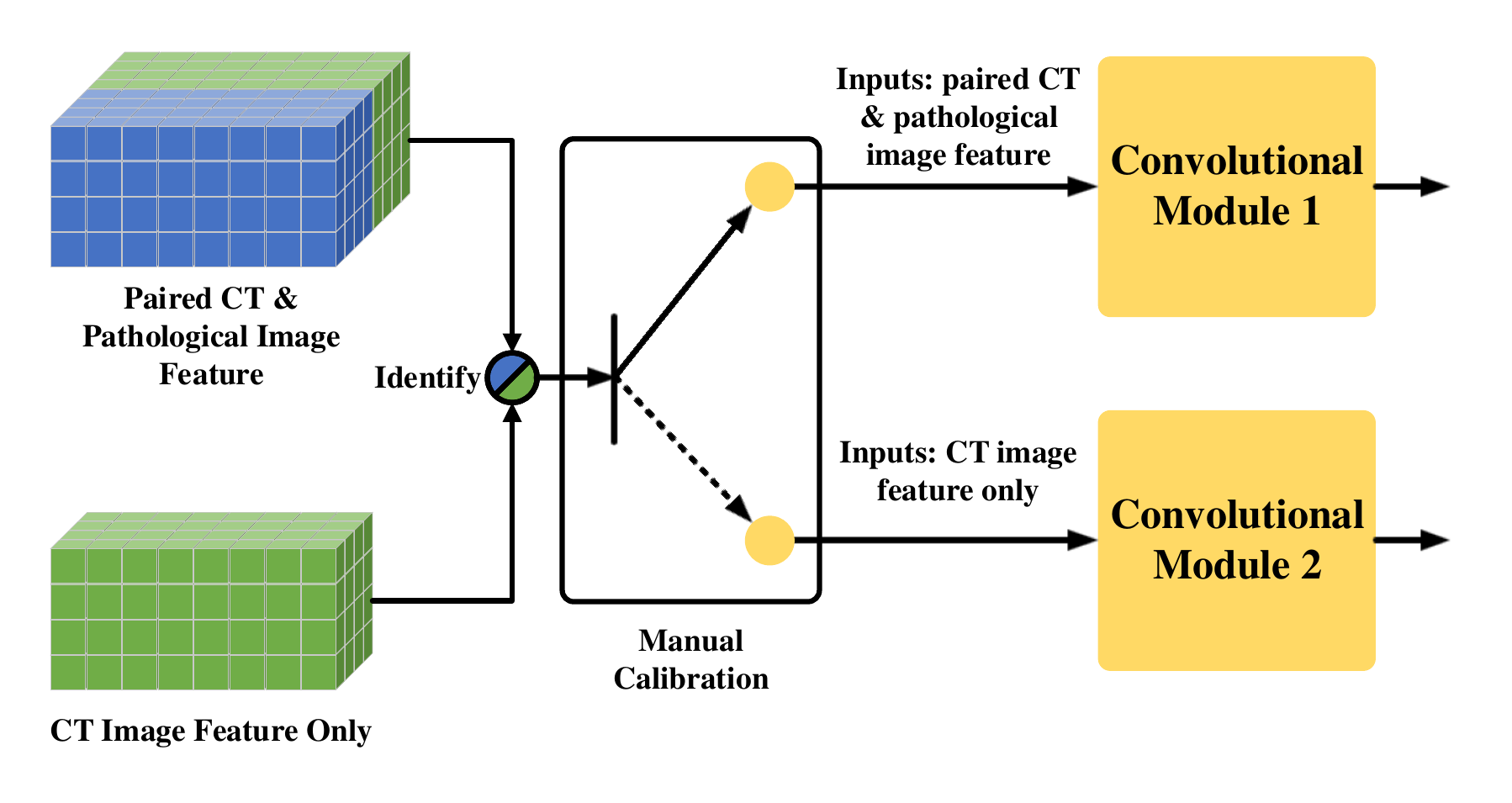}
	\caption{Key part of SwitchNet (Our proposed additional benchmark model).}
	\label{fig:SwitchNet}
\end{figure}

For the former comparative study, we compared our proposed model with other SOTAs, including ResNet-18 \cite{rs2}, ResNet-50, ViT \cite{math1}, Swin \cite{rs5}, T2T \cite{rs6}, RAFENet \cite{rs3} and CARL \cite{rs4}. Moreover, we devised an additional benchmark model SwitchNet, which allows manual calibration for the classification of cancer subtypes utilizing paired CT/pathological images or standalone CT images. The graphical representation of the SwitchNet model is shown in Fig. \ref{fig:SwitchNet}. To make a fair comparison, the feature extraction blocks of SwitchNet is kept the same as our model utilizing ViT. In this test, our model and SwitchNet were trained on the whole dataset that covers 320 paired CT/pathological images and 200 independent CT images, while other SOTAs were trained with 520 cases of CT images only. And all of the models were tested with the same testing CT images. 

\begin{figure}[!h]
  \centering
  \begin{tabular}{c}
  (a)\includegraphics[width=0.93\linewidth]{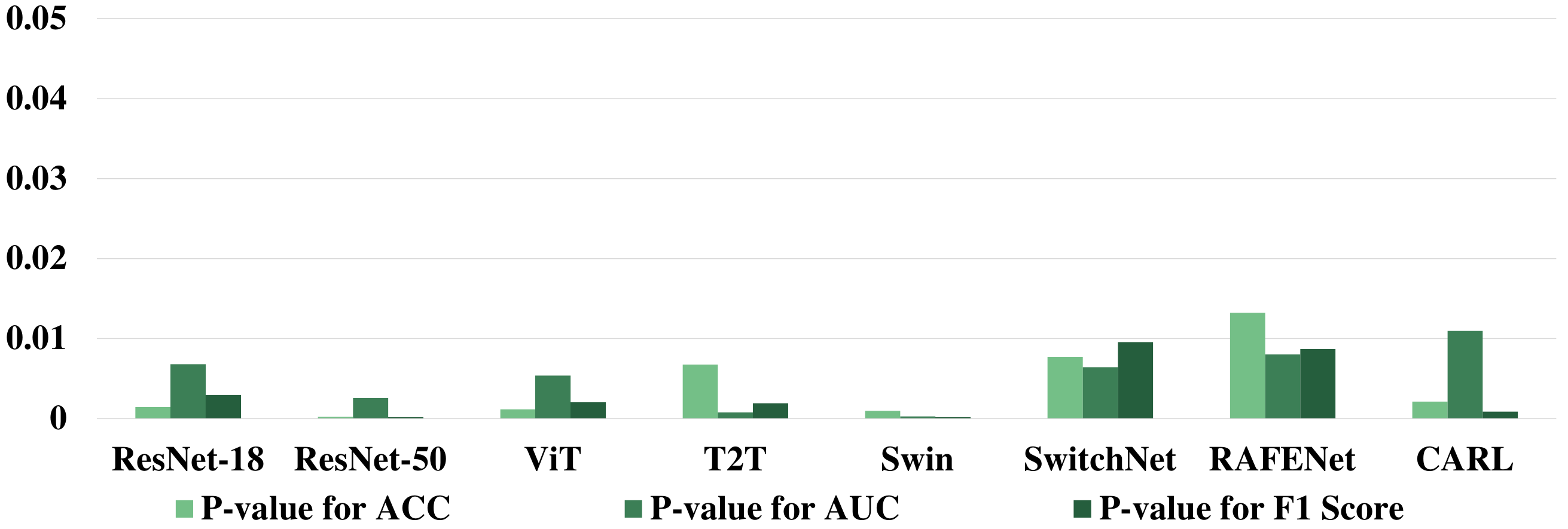} \\
  (b)\includegraphics[width=0.93\linewidth]{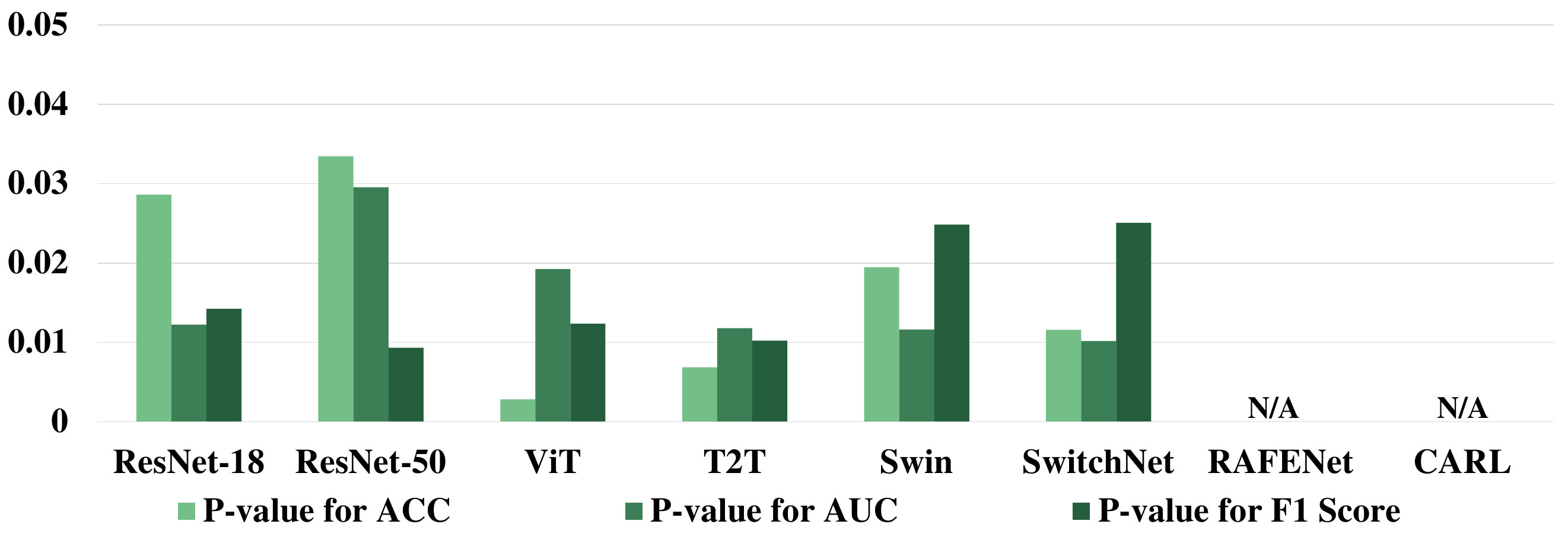} 
   \end{tabular}
  \caption{P-values obtained from paired-sample t-tests that were conducted between proposed CC-DCNet and other SOTAs for ACC, AUC and F1 SCore: (a) validated with testing CT inputs, (b) validated with testing paired CT/pathological inputs (the tests of RAFENet and CARL were skipped). All p-values are less than the significance level (0.05).}
  \label{fig:ttest}
\end{figure}
For the latter comparison study against SOTAs on paired CT/pathological dataset, we initially constructed the framework that comprises two same SOTA networks in parallel to process pathological and CT datasets, accordingly. These two networks were concatenated together at the final layer to output the predictions. In this test, the proposed CC-DCNet and SwitchNet were still trained on the whole dataset (320 paired CT/pathological images and 200 independent CT images), while other SOTAs were trained with 320 paired CT/pathological images. All of the comparative models were tested with testing CT/pathological image sets. Here, the comparison tests of RAFENet and CARL were skipped as they can only process single modality images. 


In Table \ref{tab:comparisonstudy1} and Table \ref{tab:comparisonstudy2}, the quantitative evaluation results of our model as well as other SOTAs are summarized. It may be observed that our proposed model outperforms all of the other classification methods across all metrics in both comparisons. In the first comparison with independent CT inputs, our approach achieves the results of 86.69\%$\pm$3.89\%, 90.75\%$\pm$2.51\% and 85.96\%$\pm$3.51\% for ACC, AUC, and F1-score, respectively. While the highest value of SOTAs is provided by CARL, with ACC, AUC and F1-score of 83.14\%$\pm$3.25\%, 87.61\%$\pm$3.46\% and 82.55\%$\pm$3.19\%. These numerical results reveal that introducing the relevant pathology-prior knowledge of corresponding pathological images into the radiological features (RF) based benchmark networks can significantly improve the models' accuracy. In the latter comparison with paired CT/pathological inputs, our approach achieves ACC of 96.28\%$\pm$2.56\%, AUC of 97.53\%$\pm$2.13\% and F1-score of 96.04\%$\pm$3.33\%, surpassing the highest values among state-of-the-art models obtained by SwitchNet, which are 94.65\%$\pm$2.36\%, 96.53\%$\pm$2.07\% and 94.61\%$\pm$2.98\% for ACC, AUC, and F1-score, respectively. Compared with the CT testing dataset, the performance differences among various models tested on paired multi-modality data are less obvious. Nevertheless, our model still outperforms the other models, indicating that incorporating additional independent CT data as well as setting cross-modality constraints can also improve the model's performance. 

Furthermore, we conducted paired-sample t-tests between the proposed model and other SOTA models, to assess the statistically significant differences between these methods. It can be concluded from Fig. \ref{fig:ttest} that there are statistically significant differences between the results obtained by the different methods as all p-values are less than the significance level (0.05).

\subsection{Ablation Study}
\label{subsec:ablation study}
In our model design, we construct a dynamic convolutional module and a contrastive constraint module to enable the model to adaptively process data of varying sizes, and leverage the limited pathological priors to promote the model's overall accuracy. In order to assess the impact of different input dataset on model performance and evaluate the contributions of contrastive constraints to performance enhancement, the ablation tests were carried out in this subsection.

\subsubsection{Ablation test of independent CT dataset}
A key design of our model in this study is the dynamic convolution module. This module grants our model the ability to analyze both paired CT/pathological data and independent CT data. Previous tests have confirmed that the introduction of pathological images is beneficial to the model's performance. In order to validate the impact of additional independent CT data, we further trained the proposed CC-DCNet with only paired dataset (320 paired CT/pathological images), and tested the model by testing CT data and testing paired data, separately. As shown in Fig. \ref{fig:ablation}, it can be seen that the inclusion of additional CT images also contributes to enhancing model performance, indicating that utilizing more datasets (which may not be paired) may also enhance the performance of the model. While the improvements are relatively small, which could potentially be attributed to the limited number of supplemental CT images, and further validation could be pursued by incorporating a larger quantity of CT data to offer a more comprehensive assessment of the model's performance.
\begin{table*}[!htbp]
\centering
\caption{External validations of proposed model and other comparative models on dataset from hospital B and C in terms of ACC(\%), AUC(\%) and F1-Score(\%). CT means Independent CT imgaes, CT/PI means paired CT and pathological images.}
\renewcommand{\arraystretch}{1.3}
\small
\tabcolsep=0.08cm
\begin{tabular}{cccc|ccc|ccc}
\hline 
Dataset & \multicolumn{6}{c}{Hospital B} & \multicolumn{3}{c}{Hospital C}  \\ \hline 
Inputed data   & \multicolumn{3}{c}{CT} & \multicolumn{3}{c}{CT/PI} & \multicolumn{3}{c}{CT}  \\ \hline
Metrics & ACC & AUC & F1 & ACC & AUC & F1 & ACC & AUC & F1 \\ \hline 
ResNet-18 & 75.54$\pm$3.14 & 79.00$\pm$4.02 & 73.53$\pm$3.67 & 87.91$\pm$1.75 & 91.41$\pm$1.77 & 86.98$\pm$2.25 & 71.49$\pm$2.94 & 74.17$\pm$5.88 & 70.72$\pm$5.02\\
ResNet-50 & 76.18$\pm$2.78 & 80.49$\pm$2.69 & 75.90$\pm$2.90 & 88.14$\pm$3.72 & 92.06$\pm$4.35 & 87.76$\pm$3.81 & 72.43$\pm$2.66	& 74.75$\pm$4.85 & 71.46$\pm$3.59\\
ViT & 76.49$\pm$3.67 & 80.64$\pm$4.59 & 76.11$\pm$3.17 & 88.82$\pm$2.63 & 92.08$\pm$2.35 & 88.50$\pm$2.55 &72.52$\pm$2.24 & 74.50$\pm$3.99	& 72.07$\pm$1.49\\
T2T & 76.67$\pm$3.49 & 80.60$\pm$4.36 & 76.18$\pm$3.40 & 88.53$\pm$2.51 & 91.73$\pm$1.98 & 87.89$\pm$2.66 &72.89$\pm$2.41 & 75.86$\pm$3.84	& 72.12$\pm$2.36\\
Swin & 77.01$\pm$2.55 & 80.85$\pm$3.50 & 76.32$\pm$2.81 & 89.40$\pm$2.65 & 92.59$\pm$2.29 & 88.32$\pm$2.61 &72.93$\pm$2.35 & 74.84$\pm$3.31	& 72.23$\pm$2.04\\
SwitchNet & 78.02$\pm$4.10 & 82.98$\pm$4.35 & 77.64$\pm$3.98 & 90.18$\pm$2.67 & 93.25$\pm$2.63 & 89.81$\pm$3.84  & 74.03$\pm$3.08 &76.17$\pm$4.48 &73.37$\pm$2.85\\
RAFENet & 79.08$\pm$3.16 & 82.69$\pm$3.51 & 78.51$\pm$3.18 & - & - & - &74.74$\pm$2.84 & 78.38$\pm$3.98 & 74.17$\pm$2.91\\
CARL  	& 79.79$\pm$3.91 & 83.62$\pm$4.19 & 78.81$\pm$4.09	& -	& -	& -	& 74.84$\pm$2.61 & 77.96$\pm$3.93 & 73.68$\pm$2.78\\
CC-DCNet & \textcolor[RGB]{255,0,0}{82.22$\pm$3.39}	& \textcolor[RGB]{255,0,0}{86.49$\pm$2.69} & \textcolor[RGB]{255,0,0}{81.78$\pm$2.57} & \textcolor[RGB]{255,0,0}{91.19$\pm$1.61} & \textcolor[RGB]{255,0,0}{94.98$\pm$1.74} & \textcolor[RGB]{255,0,0}{90.97$\pm$1.39} & \textcolor[RGB]{255,0,0}{77.70$\pm$3.57} & \textcolor[RGB]{255,0,0}{80.74$\pm$6.41} & \textcolor[RGB]{255,0,0}{75.72$\pm$5.74}\\ \hline
\end{tabular}
\label{tab:validation}
\end{table*}
\subsubsection{Ablation test of contrastive constraints}
\begin{figure}[!b]
  \centering
  \begin{tabular}{c}
   (a)\includegraphics[width=0.95\linewidth]{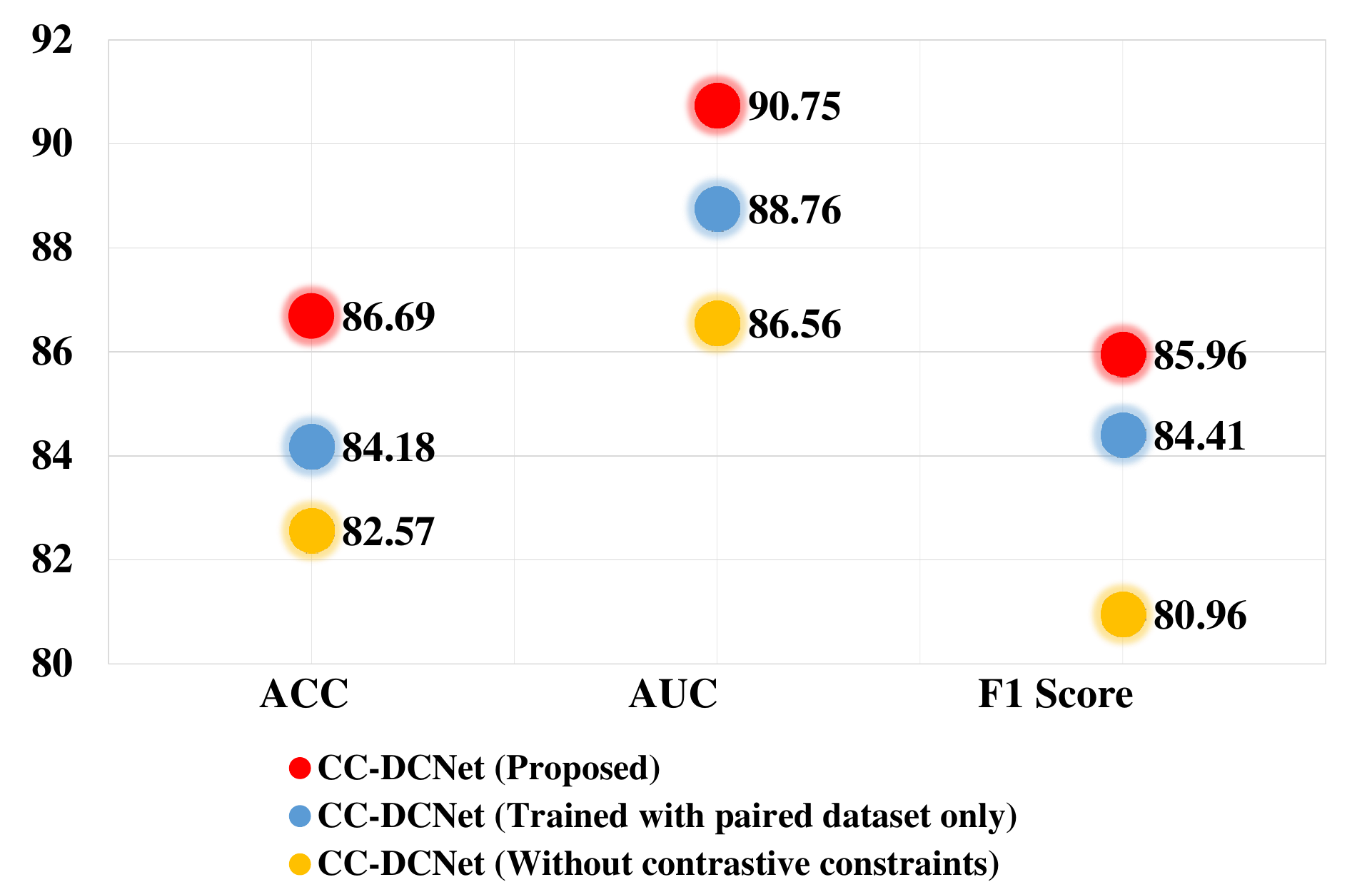}\\
   (b)\includegraphics[width=0.95\linewidth]{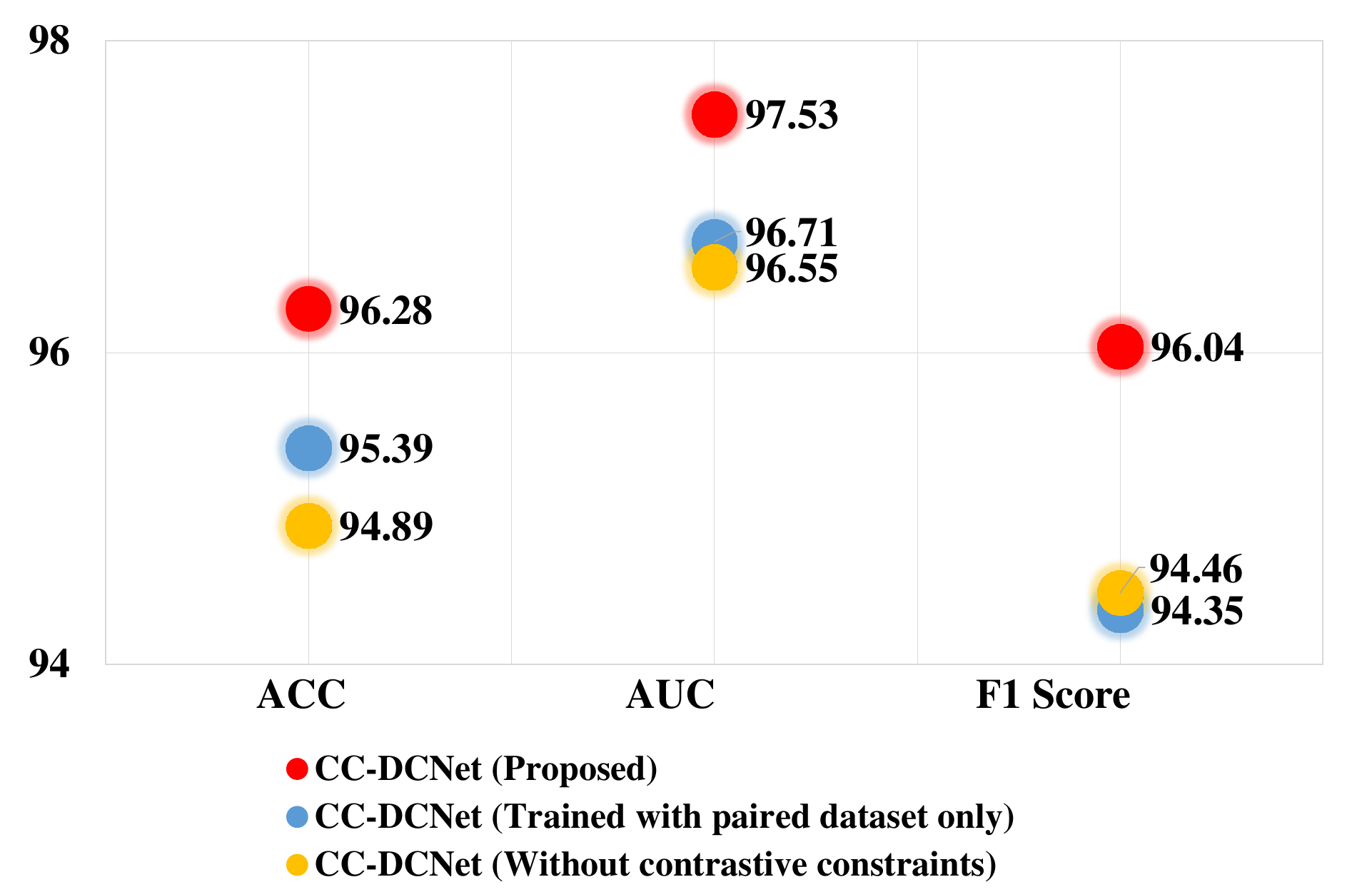}
   \end{tabular}
   \caption{The ablation test of additional independent CT dataset and contrastive constraints in terms of ACC(\%), AUC(\%) and F1-Score(\%), validated with testing CT input
 (a) and paired CT/pathological input (b).}
	\label{fig:ablation}
\end{figure}

Within the proposed model, we have designed the contrastive constraints to acquire the cross-modality correlations of paired CT and pathological images, and subsequently employ such prior information to optimize the radiological feature extractions. To investigate the contributions of contrastive constraints to model's performance, we trained the proposed CC-DCNet by removing its contrastive loss, and tested the model with the same testing datasets. In Fig. \ref{fig:ablation}, ACC, AUC, and F1-score are applied for comparisons between CC-DCNet with and without contrastive constraints. Through the comparisons, it is obvious that the model equipped with contrastive constraints demonstrates superior performance, particularly when applied on independent CT images in the test. When testing on independent CT images alone, although the paired pathological images are absent for joint analysis, the CC-DCNet with contrastive constraints is still capable of using the previously learned pathological priors to guide the feature extractions from CT images more pathologically relevant under the constraints of contrastive loss, and finally produces a more accurate diagnosis. 
\subsection{Impact Analysis of Feature Extractor}
\begin{figure}[!t]
  \centering
  \begin{tabular}{c}
 \includegraphics[width=\linewidth]{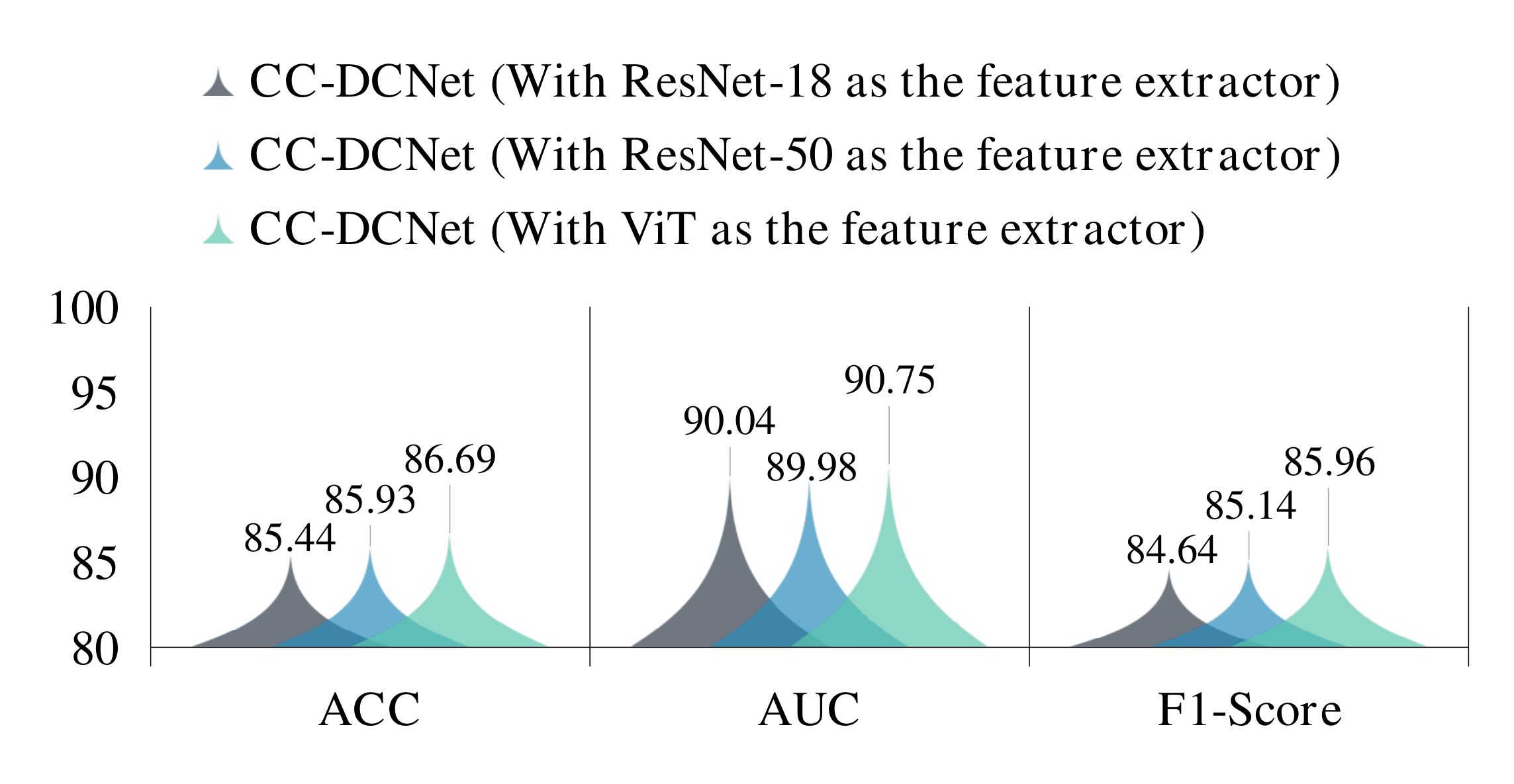}\\(a)\\
 \includegraphics[width=\linewidth]{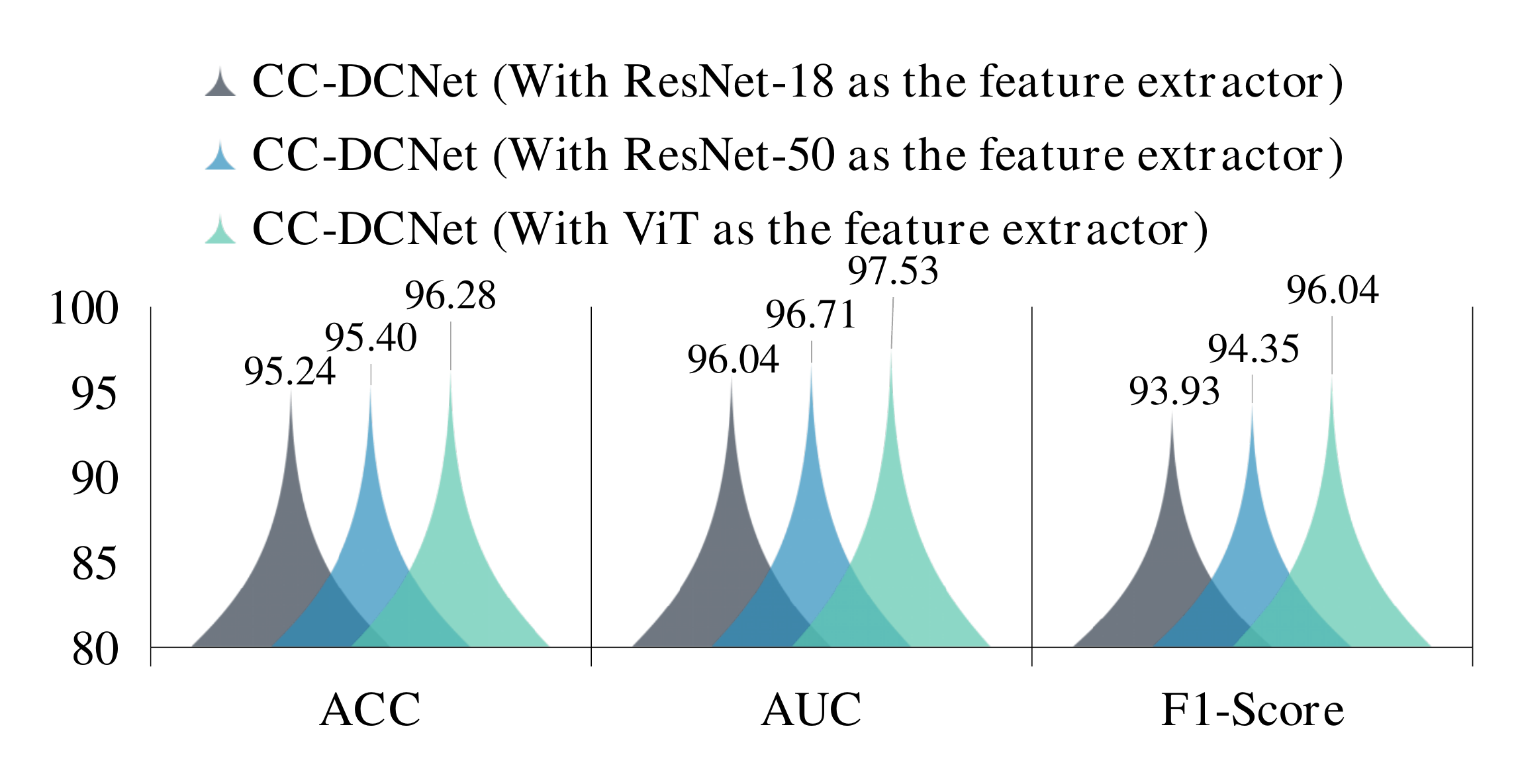} \\(b)
   \end{tabular}
  \caption{The impact analysis of feature extractor, which is equipped with ResNet-18 (CASE A), ResNet-50 (CASE B), ViT (CASE C), in terms of ACC(\%), AUC(\%) and F1-Score(\%) with validating CT inputs (a) and paired CT/pathological inputs (b).}
  \label{fig:impact_analysis}
\end{figure}

In this work, we applied ViT as the feature extraction blocks for CT and pathological images. To investigate the impact of different network structures on the model's performance, we further trained and tested the proposed CC-DCNet with ResNet-18 and ReNet-50 as feature extractors. The number of parameters remain the same across different comparative models. As presented in Fig. \ref{fig:impact_analysis}, the model constructed with ViT outperforms the other constructions on both testing CT data and testing paired data, revealing that the attention mechanism in ViT is advantageous for extracting valid features from large-size images.

\subsection{External test}
\label{subsec:external validation}
This study was conducted using an extensive multi-center dataset that includes data from three distinct hospitals. By utilizing the training dataset from Hospital A, we have effectively demonstrated the advancements of our proposed model over other models through a series of experiments. In this subsection, we further validate our well-trained model on two additional datasets from Hospital B and Hospital C, to assess the model's generalization ability and robustness. It is worth mentioning that we collected paired CT/pathological datasets from Hospital B while independent CT images from Hospital C. 

Table \ref{tab:validation} presents the outcomes of external performance validations, drawing comparisons between our proposed model and other SOTAs. However, it has to be acknowledged that both the comparison models and our model exhibit relatively diminished performance on data from Hospital B and Hospital C as compared to that from Hospital A, they still yield reliable levels of ACC, AUC and F1-score. And of paramount significance is that when tested on the same dataset, our proposed classification model consistently outperforms other models, and these improvements are substantial. The proposed model achieves a minimum of 2.43\%, 2.36\% and 1.55\% improvements accordingly in ACC, AUC, and F1 score for independent CT inputs, and 1.01\%, 1.73\%, and 1.16\% in ACC, AUC and F1 score for paired CT/pathological inputs.

\section{Discussion}
\label{sec:discussion}
Early identification of pathological subtypes is of significant importance for lung cancer patients. In clinical practice, CT imaging stands as one of the most frequently employed techniques for diagnosis. Consequently, there is a huge demand for CT-based automated diagnostic models. However, the key challenge in utilizing such CT models for predicting pathological subtypes lies in the issue of accuracy. In cancer diagnosis, pathological examination serves as the gold standard, providing more distinct and precise diagnostic criteria. Therefore, in this work, a significant hypothesis is that the integration of pathological information into the fundamental CT-based model, forming the comprehensive multi-modality features, holds the potential to enhance the classification accuracy of lung cancer pathological subtypes. While, it should be noted that the invasive pathological examinations may not be applicable to all clinical scenarios, which would result in missing modality and consequently the presence of a substantial amount of paired multi-modality as well as single-modality dataset. To enhance the overall utility of the clinical dataset and acquire prior information on cross-modality correlations from paired datasets, guiding the model to make more accurate predictions even in the absence of pathology, we proposed a dynamic convolutional neural network with contrastive constraints specifically designed for identifying lung cancer subtypes with CT and pathological images.

Across various homologous but diverse modality images of the same patient, mutual characteristics of the same disease may be revealed from different perspectives, thereby the integration of multi-modality information holds the superior diagnostic evidence. Comparing the results in Table \ref{tab:comparisonstudy1} and Table \ref{tab:comparisonstudy2} with either independent CT inputs or paired CT/pathological inputs, our hypothesis may be confirmed where the incorporation of pathological information significantly enhances the CT-based model’s classification accuracy compared to models that are trained exclusively with CT dataset only. 

Generally, the common issue of modality missing in clinical practice poses a big challenge to conventional models. In this work, in order to empower the model to deal with diverse input combinations and further enhance the applicability and representation capacity of clinical computer-aided systems, a dynamic convolution module was designed. It allows the model to adaptively adjust itself for various inputs through the multiple parallel convolution kernels. Through the dynamic convolution module, the proposed model achieves training simultaneously with either independent CT images or paired CT/pathological images as inputs, thereby offering the potential for a comprehensive analysis of various data combinations in clinical scenarios. For the construction of the proposed model, considering the computing limitations of GPU, a transformer-based approach was adopted to extract features from the CT and pathological images by patch-wise sampling. Transformer network can process the data in all positions of the input sequence in parallel, and efficiently capture the long-range dependencies in input sequence through the attention mechanism. As shown in Fig. \ref{fig:impact_analysis}, using Vision Transformer (ViT) as the feature extraction blocks resulted in performance improvements compared to the ones using ResNet as a feature extractor, indicating that the attention mechanism in ViT is beneficial for extracting relevant features from large-size images. In addition, given that the spatial positioning of patches in WSI has minimal impact on the final pathological representation, whereas information related to the magnification level effectively signifies the level at which the patch represents pathological information, the original position encoding was replaced with magnification encoding to better extract features from pathological images at varying magnification levels. 

Despite the substantial differences between CT and pathological images, several literature have shown the presence of cross-scale correlations between these two modalities, e.g., the specific relationships between CT intensity values and matched cell density statistics \cite{math3}, although investigation on deep-level correlation is still far from enough. Therefore, to extract more pathological-relevant features from CT images and provide pathology prior guidance even when paired with pathological data is missing, this work incorporated a contrastive loss constraint into the framework of the DC network. From the results in Fig. \ref{fig:ablation}, it may be seen that in the absence of contrastive constraint, the model trained with paired CT/pathological images maintained its performance, however, it showed a substantial drop in performance when it was trained solely on CT images. This indicates that the contrastive learning module plays a relatively crucial role in leveraging pathology priors to aid CT analysis. The introduction of pathology information has been proved to be pivotal in model optimization. Moreover, to assess the impact of expanded single CT images on the model’s performance, an ablation study was conducted by keeping the model structure unchanged while training the model with only paired CT/pathological images. From the comparison results in Fig. \ref{fig:ablation}, it can be concluded that the extra independent CT data also contributed to the overall performance improvement of the model. Although the improvements are not very remarkable, which could be attributed to the limited number of extra-independent CT samples (only 200 cases), increasing the number of CT images could reveal more pronounced differences. 

Overall, our proposed model offers a more accurate diagnosing aid of lung cancer pathological subtypes, and the concept of dynamic convolution module can be applied to other multi-modality scenarios, providing a new direction for addressing complex multi-modality data challenges. There is also room for improvements in our model. According to the results in external evaluations, our model gives more accurate predictions compared to other SoTAs, while the multi-center effect still exits due to the variations in equipment or operation settings. In future work, the compensation of the multicenter effect with interpretable models will be studied, where the more interpretable index between CT and pathological images could be introduced into the model to improve its generalization ability. 

\section{Conclusion}
\label{sec:conclusion}
In this research, we introduce an innovative deep learning model designed for the classification of different subtypes of lung cancer. Our model features a combined contrastive learning module that leverages pathological features as prior knowledge to enhance performance even when using standalone CT input. Additionally, we have incorporated a dynamic convolution module capable of generating convolution kernels tailored to input features of varying dimensions. This empowers our model to seamlessly handle both paired CT/pathological images and standalone CT images. Extensive experimentation conducted on a substantial multi-center dataset highlights the superior performance of our model compared to other state-of-the-art methods. This superiority is evident through improved metrics such as ACC, AUC, and F1-score results. Furthermore, the strategic integration of pathological priors, as proposed in our approach, holds significant potential for extension into even more intricate and challenging applications.

\end{document}